\documentclass[
    ,final            
  ]
  {aipproc}

\layoutstyle{8x11double}

\usepackage{natbib}

\begin{document}

\title{Space cowboys odyssey: beyond the Gould Belt}


\author{Popov S.B.}{
  address={Sternberg Astronomical Institute,
Universitetski pr. 13, 119991 Moscow, Russia}
}

\author{Posselt B.}{
  address={Observatoire Astronomique de Strasbourg,
11 rue de l' Universite, 67000 Strasbourg, France}
  ,altaddress={Astrophysikalisches Institut und
Universit\"{a}ts-Sternwarte, Schillerg\"{a}\ss chen 2-3,
07745 Jena, Germany}
}

\author{Haberl F.}{
  address={Max-Planck-Institut
  f\"{u}r extraterrestrische Physik, Postfach 1312
85741 Garching, Germany}
}

\author{Tr\"umper J.}{
   address={Max-Planck-Institut
  f\"{u}r extraterrestrische Physik, Postfach 1312
85741 Garching, Germany}
}
\author{Turolla R.}{
   address={University of Padua, Department of Physics,
via Marzolo 8, 35131 Padova, Italy}
}
\author{Neuh\"auser R.}{
   address={Astrophysikalisches Institut und
Universit\"{a}ts-Sternwarte, Schillerg\"{a}\ss chen 2-3,
07745 Jena, Germany}
}

\begin{abstract}
We present our new advanced model for population
synthesis of close-by cooling NSs. 
Detailed treatment of the initial spatial
distribution of NS progenitors and a detailed ISM structure up to 3 kpc give
us an opportunity to discuss the strategy to look for new isolated cooling
NSs.
Our main results in this respect are the following: new candidates are
expected to be identified behind the Gould Belt, in directions to rich OB
associations, in particular in the  Cygnus-Cepheus region; new candidates,
on average, are expected to be hotter than the known population of cooling NS.
Besides the usual approach (looking for soft X-ray sources), the search in 
'empty' $\gamma$-ray error boxes 
or among run-away OB stars may yield new X-ray
 thermally emitting  NS candidates.
\end{abstract}


\maketitle


\section{Introduction}

More than 10 years after the discovery of its
brightest member RX J1856-3754 \cite{Walter1996},
a group of seven radio-quiet isolated neutron
stars (NSs) detected by ROSAT
gained an important place in the
rich zoo of compact objects. Together with Geminga and several close-by
young radio pulsars, these objects 
form the local population of cooling NSs.
Studies of this group of sources already provided a wealth of information on
NSs physics (see e.g.
\cite{Haberl2007,Page2007,Zane2007} for recent
reviews).

Since 2001 the number of known close-by radio-quiet NSs has not been growing
despite all attempts to identify 
new candidates.
Partly this is due to the fact that all these searches are blind.
 To advance the identification of new near-by cooling NSs it is
necessary to perform a realistic modeling of this population.

To investigate the population of close-by young cooling NSs the method of
population synthesis is used here.
In this short note  an advanced population synthesis model is briefly
discussed for the
population
of close-by ($< 3$~kpc) isolated NSs which can be
observed via their thermal emission in soft X-rays (the detailed description
and full analysis of new results will be presented elsewhere
\cite{Posselt2008}).
Previously our models were applied to confirm the link between the
seven radio quiet NSs (the Magnificent
Seven) and the
Gould Belt \citep{p03} (Paper I), to study distribution of NSs in the Galaxy
and in the solar vicinity \cite{p04} (Paper II),
and to test theories of thermal evolution of NSs \citep{pgtb04} (Paper III).
The major interest of the present study is to
get a hint how to find more objects of this type.


\section{The new model}

The main physical ingredients which constitute our population synthesis
model are:

\begin{itemize}
\item[(A)] the initial NS spatial distribution and the NS birth rate;
\item[(B)] the kick velocity distribution of the NSs;
\item[(C)] the Galactic gravitational potential;
\item[(D)] the NS mass spectrum;
\item[(E)] the NS cooling curves;
\item[(F)] the NS surface emission in X-rays;
\item[(G)] the interstellar absorption of X-rays;
\item[(H)] the properties of the X-ray detector.
\end{itemize}

Ingredients B, C, E, and F are unchanged with respect to our previous
studies.
The ingredients A, D, G, and H are modified. 
The new detailed inital spatial distribution of NS progenitor is one of the
main feature of the advanced model. Now we take into account inhomogenities
of the distribution of massive stars up to 3 kpc from the Sun (OB
associations and other stellar groups).
We use new ISM element abundances and photoelectric cross
sections from \cite{Wilms2000}, and apply two new
variants of the ISM 3D distribution.
We perform more accurate calculations
of the detector response than it was in Papers I-III. Some details about
these modifications can be found in \cite{Posselt2006, Posselt2008}.

At each time step
we consider eight NS
masses with corresponding cooling curves.
The overall result of simulated  evolutionary tracks,
from birth till the time when the temperatures falls below $10^5$~K,
 is normalised by the mass distribution as well as by birth rates.

\section{Results}

 At first, it is necessary to note that significant advances made in the new
model do not change significantly  the main results (in the first place 
the Log~N~--~Log~S distribution) in
comparison with our early studies. In particular, we tested a new variant of
NS mass spectrum in which the peak at $\sim 1.4$~$M_\odot$ is smeared out. This
modification has insignificant influence on the Log~N~--~Log~S distribution.
The use of more realistic abundances of elements in the ISM and more precise
calculation of the detector response also have minuscule effects (see
\citep{Posselt2006}).
Finally, the new initial spatial distribution of NS progenitors and two new
models of the ISM distribution have  visible, but not crucial influence on
the Log~N~--~Log~S distribution (see, Fig.~1 for the effect of the new ISM
models). With the new initial spatial distribution, cooling curves from
\cite{bgv2004} (model I in Paper III), and the new ISM model based on data
from \cite{Hakkila97} (dashed line in Fig.~1) we can successfully fit
data points at all fluxes. Our prediction about the number of unidentified
cooling NSs in the solar vicinity is generally unchanged with respect to our
previous calculations. However, now in addition to calculations of absolute
numbers of unidentified objects, we can address the question of their 3D
distribution, and their age distributions for different fluxes.


\begin{figure}[h,t]
  \includegraphics[width=0.45\textwidth]{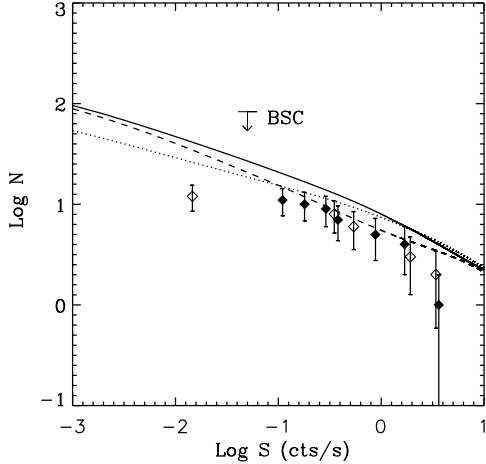}
  \caption{Log N~--~Log S for different X-ray absorbing ISM models.
All curves are plotted for the new initial spatial distribution.
Solid curve: old, simple analytical ISM model as e.g. in
Paper III.
Dotted curve: new improved analytical ISM model;
dashed curve: ISM model which includes the extinction study by
Hakkila (1997).}
\end{figure}


New results obtained in this study 
are related to artificial maps of cooling NS distributions on the
sky, and to age and distance distributions of sources in different flux
ranges. These new data give us an opportunity to discuss a strategy to
identify new candidates.

On the calculated maps
sources appear to be concentrated towards the Galactic plane and
the plane of the Gould Belt (see Fig.~2). 
Few objects are expected to be found
at latitudes higher than 30$^\circ$ (however, here we do not take into account
runaway and hypervelocity progenitors, which can be important to study such
objects as Calvera  \citep{Rutledge2007}). 
Inside $\pm$~30 degrees from the Galactic
plane the distribution of sources is dominated by NSs from
relatively close, rich OB associations
(Sco OB2, Cyg OB7, Cep OB3, and Ori OB1).

Interplay between source distribution and 3D ISM structure allows
us to make predictions on which directions are most promising for
looking for new cooling NSs. Our results indicate that isolated cooling NSs
should be searched in directions of OB associations such as Cyg
OB7 and Cep OB3. Analysis of the age distributions of sources in different
flux ranges shows, that on average, new candidates should be slightly
hotter than the Magnificent Seven  as they are younger (this is an obvious
selection effect: it is easier to detect  hotter sources from larger distance
observing through absorbing ISM).

\begin{figure}[h,t]
  \includegraphics[width=0.45\textwidth]{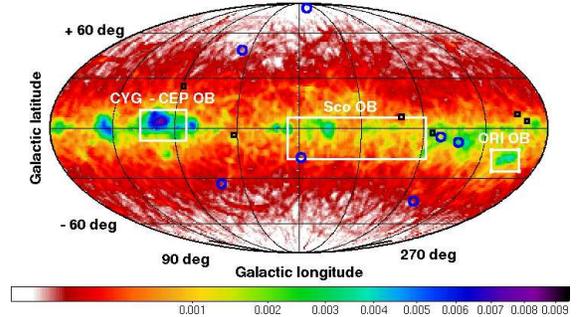}
  \caption{The expected number density of isolated neutron stars
with thermal X-ray emission in units of numbers per square degree.
The galactic map is in Mollweide projection.
Only sources with ROSAT PSPC count rates larger than 0.05~cts s$^{-1}$ are
considered, the same value as used in Paper II in their Fig.~6.
The simulation was done for new initial progenitor distribution,
new abundances, old mass spectrum and the analytical ISM model,
thus corresponds
to the dotted Log~N~--~Log~S curve in Fig.~1.
Circles mark the positions of the Magnificent Seven, and
squares -- the positions of close young radio pulsars with detected thermal
X-ray emission.}
\end{figure}

Analysing the distance distribution of sources with different fluxes, we
note the following feature. 
While for large
fluxes  (brighter 0.1 ROSAT counts per second)
most sources are situated in the region  $\sim 200$~-~$
400$~pc (ie. in the Gould Belt), for  fainter objects the picture
is different. At fluxes below $\sim 0.1$~cts~s$^{-1}$ one expects to
see mostly sources behind the Gould Belt.
These general features are very important, as they indicate that
new, still unidentified ``cowboys'' are expected to be young objects
behind the Gould Belt.

\section{Discussion}

The main aim of this study is to make some advances in the
strategy for searching for new isolated cooling NSs. 
According to our results, 
new candidates expected to be identified at
ROSAT count rates $<0.1$~cts~s$^{-1}$ should be young objects born
in rich OB associations behind the Gould Belt. Most of the recent
studies \cite{Agueros2006,Chieregato2005,Rutledge2003}
looked for new candidates far from the galactic plane. It seems
that this is not very promising. Our results indicate that new cooling NSs
should be searched in directions of OB associations such as Cyg
OB7 and Cep OB3. 

Considering sky coverage the ROSAT All Sky Survey is currently the
best choice to look for new ``cowboys'' in the Cygnus-Cepheus region which is,
according to our results, the most promising area. However, the
relatively large positional error circle of ROSAT usually includes many
possible optical counterparts, especially at these low galactic latitudes.
Furthermore one has to exclude variable X-ray sources to find isolated
cooling NSs. In
this respect the recently published XMM-$Newton$ Slew Survey 
 may become an important
database. A major step can be expected from the planned eROSITA all sky
survey which - compared with the ROSAT all sky survey --  will provide a
factor of $\sim$~10 in soft X-ray sensitivity and factor of $\sim$~4 in energy
resolution \cite{Predehl2006}.


 Some of unidentified $\gamma$-ray sources (already observed by EGRET and
forthcoming due to AGILE and GLAST) can be identified as
cooling NSs as it was with Geminga and 3EG J1835+5918. 
In particular, GLAST observations of the 56 EGRET error boxes studied in
\cite{Crawford2006} and later cross-correlation with the ROSAT (or/and XMM)
data can result in new identification of cooling NSs.

Another possibility to find new isolated coolers
 is to search for (un)bound compact
companions of OB runaway stars. More than one hundred OB runaway stars are
known in a 1 kpc region around the Sun \citep{Zeeuw1999}. 
They are characterized by large spatial velocities or/and
by large shifts from the galactic plane. Two main origins of these large
velocities are currently discussed: dynamical interaction and explosion of a
companion in a close binary system. The latter case is interesting for the
discussion of search for new close-by cooling NSs.

A binary can survive after the first SN explosion in, roughly,
10-20\% of cases. Then one expects to have a runaway system
consisting of an OB star and a compact object (most probably a
NS). A young NS can appear as a radio pulsar. In \cite{sayer1996} and
\cite{philp1996} the authors
searched for radio pulsar companions of $\sim 40$
runaway OB stars. Nothing was found. This result is consistent
with the assumption that in less than 20\% cases OB stars have
radio pulsar companions. Still, it is interesting to speculate
that runaway massive stars can have cooling radio quiet NS as companions. Then a
companion can be identified as a source of additional X-ray
emission.

\begin{theacknowledgments}
We thank D. Blaschke, H. Grigorian, and D. Voskresensky for data on cooling
curves and discussions; A. Mel'nik for discussion of properties of OB
associations; R. Lallement for the sodium data;
and A. Pires for discussions about the ISM model.
S.B.P. was supported by INTAS and Dynasty foundations.
\end{theacknowledgments}


\bibliographystyle{aipproc}   


\bibliography{mont}


\end{document}